# Temperature dependent dynamic and static magnetic response in magnetic tunnel junctions with Permalloy layers

J. F. Sierra, V.V.Pryadun and F.G.Aliev[*]

*Dpto. Física de la Materia Condensada, C-III, Universidad Autónoma de Madrid, 28049, Madrid, Spain.*

S. E. Russek

*National Institute of Standards and Technology, Boulder, CO 80305, USA.*

M. García-Hernández
*Instituto de Ciencia de Materiales Madrid, CSIC, Cantoblanco, 28049, Madrid, Spain*

E. Snoeck
*CEMES- CNRS, Toulouse, France*

V. V. Metlushko
*Dept. Electrical and Computer Engineering, University of Illinois at Chicago, Chicago, Illinois, USA*

REVISED VERSION: September 25-2008

## Abstract

Ferromagnetic resonance and static magnetic properties of $CoFe/Al_2O_3/CoFe/Py$ and $CoFe/Al_2O_3/CoFeB/Py$ magnetic tunnel junctions and of 25nm thick single-layer Permalloy (Py) films have been studied as a function of temperature down to 2K. The temperature dependence of the ferromagnetic resonance excited in the Py layers in magnetic tunnel junctions shows "knee-like" enhancement of the resonance frequency accompanied by an anomaly in the magnetization near 60K. We attribute the anomalous static and dynamic magnetic response at low temperatures to interface stress induced magnetic reorientation transition at the Py interface which could be influenced by dipolar soft-hard layer coupling through the $Al_2O_3$ barrier.

Since the discovery of large room temperature magnetoresistance in magnetic tunnel junctions (MTJs)[1,2] there have been continuous efforts to improve the quality of MTJs and optimise their dynamic response, important for different applications.[3-5] In order to optimize the low field performance, the MTJ sensors with amorphous $Al_2O_3$ barriers employ magnetically soft ferromagnetic (FM) films typically made of Permalloy ($Ni_{80}Fe_{20}$). Knowledge of the dynamic and static magnetic response of Py in a wide temperature range is therefore crucial in a view of potential low temperature applications of these spintronic devices. It was observed in the late 1960's that the temperature dependence of the ferromagnetic resonance (FMR) for in-plane magnetized single Py films shows anomalous variation of the FMR linewidth[6] which has been attributed to spin-impurity interaction enhanced damping. More recently, enhancement of the FMR frequency in single Py films with decreasing temperature has been observed[7] and explained in terms of spin reorientation transition at the Py interface below 100K. This Letter presents an investigation of the temperature dependence (2K<T<300K) of the dynamic and static magnetic properties of about 25nm single-layer Py films and of Py films inserted as a free layer in MTJs with $Al_2O_3$ tunneling barriers. Our results indicate a magnetic reorientation transition (RT) in Py films at low temperatures (T<$T_R$). The static and dynamic response in MTJs could be affected by soft-hard layer dipolar coupling and interface stress in Py films.

The 25nm single-layer Py film (sample A) was grown by electron-beam evaporation on standard Si(100) wafer covered by a 2.5nm naturally-oxidized $SiO_2$ layer without an external magnetic field applied during deposition and without a top capping layer. In addition, two different MTJs optimised for low field sensors applications[8] were grown in a high vacuum sputtering chamber. Both MTJs had the following structure: Pinned-FM/Al$_2$O$_3$(1.8nm)/Free-FM with Pinned-FM being exchanged biased pinned magnetic electrode composed of Ir$_{20}$Mn$_{80}$(10nm)/Co$_{90}$Fe$_{10}$(3nm). The free ferromagnetic electrodes (Free-FM) were Co$_{60}$Fe$_{20}$B$_{20}$(2nm)/Ni$_{80}$Fe$_{20}$(23nm) for sample B and Co$_{90}$Fe$_{10}$(3nm)/Ni$_{80}$Fe$_{20}$(28nm) for sample C. The entire wafers were covered with a Ta(5nm)/Cu(5nm) layers to prevent oxidation and were annealed at 250ºC in the sputtering chamber during one hour in an in-plane applied magnetic field of 20mT. For more details on sample growth and characterization see references.[8,9]

The low temperature FMR experiments were carried out with a commercial Agilent Vector Network Analyzer (VNA) working up to 8.5GHz by employing VNA-FMR technique[10] which uses a coplanar wave guide to create the pumping field h$_{RF}$. For cryogenic measurements (2K<T<300K) we used a variable temperature cryostat with an RF insert. In order to determine the dynamic response of the Free-FM layer, an in-plane external bias field H$_{ap}$ was applied along the easy axis with h$_{RF}$ being transverse to H$_{ap}$. The data analysis is described in more details in Ref.10. The resonance frequency f$_0$ and linewidth Δf$_0$ were evaluated by fitting the resonance peaks to a lorentzian curve. A Quantum Design SQUID magnetometer was used to measure the field dependence of the in-plane magnetization M up to 0.5T at different fixed temperatures. In addition, the structures of the different layers and interfaces were investigated at room temperature by transmission electron microscopy (TEM) in high resolution mode (HRTEM) using a FEI-F20 microscope with point resolution of 0.12 nm.

Figure 1 shows the loss profile (imaginary part of microwave permeability parameter U(f), see Ref.10) and compares the magnetic field dependence of the FMR frequency measured at 150K and at 10K for the three samples investigated. In order to describe in more detail the temperature dependence of FMR, Fig. 2 compares $f_0(T)$ and $\Delta f_0(T)$ for samples A-C determined with field $\mu_0 H_{ap}$=20mT. Both MTJ samples (B and C) show independent of the magnetic field history an increase of the FMR frequency below a temperature $T_R$ of about 65K, with the increase being more pronounced for the MTJ-C (see also Fig. 1). These changes in the FMR are accompanied by nearly step-like changes in the FMR linewidth below $T_R$ for both MTJs. The single Py film (sample A), however, shows a more gradual enhancement of the FMR frequency and the linewidth below roughly 100K, which is in a good agreement with previous studies.[6,7] The presence of a weak step-like increase in the FMR linewidth just below 55K in sample A can also be observed in Fig. 2b.

Figure 3a illustrates quasi-static magnetization curves for the MTJ-C measured for three temperatures close to 60K. Here we shall focus on the field intervals where both Free-FM and Pinned-FM layers are forced to the mostly parallel alignment (P) and on the region of the antiparallel alignment (AP). By analyzing the low field hysteresis loops of the Free-FM layer near P=>AP transition we determine the switching field of the Free-FM layer as a function of temperature (Fig. 3b). One clearly sees that in both types of the MTJs studied the switching field is substantially lower than observed in the single Py film. This difference may indicate the *presence of dipolar fields induced interaction between hard and soft layers in magnetic tunnel junctions* which influences domain walls nucleation and propagation in the soft layer.

For all samples the high field magnetization curves M(H) in the P state show magnetization saturation above few 0.2T, except in the proximity to 55-70K where some deviation from the saturation behavior with an anomalous peak and dip in the M(H) dependence below and above a critical temperature is observed. Data for the MTJ-C, where this unusual behavior in M(T,H) in the P-state is mostly remarkable, with up to 20% deviation in the high field magnetization, is shown in Fig. 3a. In order to compare M(H,T) dependences for samples A-C, Fig. 3c plots the temperature dependence of the high field magnetization in the applied field of 0.5T (further $M^*_S$) normalized by the corresponding $M^*_S(5K)$ values. For the all samples studied $M^*_S(T)/M^*_S(5K)$ is close to one, except in proximity to the temperature interval around 55-70K where an anomaly is observed. One clearly sees that this magnetization anomaly is strongest and shows *qualitatively different temperature dependence* with peak and dip close to 60K for the MTJ samples B,C in comparison with the single Py film (A) where only weak maximum (about 3% deviation) is seen (see inset of Fig. 3c) which may reflect transition from perpendicularly oriented Py interface spins at $T \ll T_R$ to in-plane disordered Py interface spins at $T \gg T_R$.

We believe that the high field magnetization anomaly is evidence of a magnetic reorientation transition roughly below $T_R \approx 60K$ in the magnetically soft layer, supporting previously described FMR vs. temperature results. To show this more clearly, we marked with arrows the positions of the corresponding anomalies in $M^*_S(T)$ in Fig. 2a, showing FMR(T). The differences in the temperature dependence of FMR and magnetization in MTJ-B, C compared to single Py film (A), may indicate some fundamental changes in the RT occurring in the soft layer, most probably related to the presence or absence of magnetically hard layer.

In order to understand these differences in the reorientation transition which include the anomalous behavior in the $M^*_S(T)$ close to $T_R$ accompanied by clear "knee-like" variation of the FMR frequency in MTJs B and C, we propose a simple model (see sketch in Fig. 4) which considers the mutual influence of the pinned (hard) layer and the free (soft) one, induced by dipolar antiferromagnetic coupling through the $Al_2O_3$ barrier with some anticorrelated roughness.[11] Figure 4a sketches magnetization in the regions of the soft and hard layers interfacing $Al_2O_3$ barrier at temperatures $T \ll T_R$, when the MTJ stack is situated in 0.5T field. These layers may have magnetic moments directed mostly out-of-plane due to the surface anisotropy in the Py and dipolar coupling. This means that both interface magnetic moments tend to occupy a relative minimum of the energy corresponding to an out-of-plane magnetization (see sketch for the related energy profiles in Figs. 4a-d). Here "springs" indicate dipolar coupling between the hard and the soft layers sketched by "balls". Close to $T_R$; at $T \leq T_R$ (Fig. 4b) the soft layer starts to undergo an orientational transition from the out-to-plane to in-plane alignment enhancing the in-plane magnetization $M^*_S(T)$, by moving the soft ferromagnetic system (Py) toward its global energy minimum. Figure 4c schematically shows what may happen at $T \geq T_R$ when the hard layer, due to its coupling to the soft layer and due to the strong difference in the soft and hard layer metastable energy profiles, is pushed towards in-plane (trending to be antiparallel to the soft layer) magnetization configuration, reducing therefore the total in-plane magnetization $M^*_S(T)$. Finally, at $T \gg T_R$ (Fig. 4d) both soft and hard layers turn to have in-plane magnetization (both "balls" in the sketch in Fig. 4d occupy their absolute minima of energy) in equilibrium conditions with an antiparallel alignment showing a total magnetization of the stack nearly the same as at $T \to 0$. Within our model, the quantitative differences between response observed in MTJs samples B and C could be attributed to different materials interfacing the Py layer (CoFe or CoFeB) which could determine the stress at the Py interface.

*In Conclusion*, temperature dependent dynamic and static magnetic response in magnetic tunnel junctions with Permalloy layers shows a magnetic reorientation transition below 60K which is qualitatively different from one reported for single Py films,[7] most probably due to dipolar soft-hard layer coupling. These findings could be important for low temperature applications of devices incorporating Permalloy.[12]

**Acknowledgements**

We thank Referee for suggesting soft-hard layer coupling mechanism to describe temperature dependent high field magnetization. Discussion with A.Levanuyk and R.Heindl is gratefully acknowledged. Authors acknowledge support by Spanish MEC (MAT2006-07196; MAT2006-28183-E; MAT-2005-06024-C02-01) and U.S. NSF (ECCS-0823813 (VM)).

**FIGURE CAPTIONS**

FIG. 1.(Color on-line): Contour plots of the FMR at different external bias fields. The color scale indicates the magnitude of the imaginary part of the susceptibility (see Ref. 10). The top panel (a-c) shows resonant frequency $f_0(H)$ measured at 150 K while the bottom panels (d-f) show $f_0(H)$ at T=10K.

FIG. 2. (Color on-line) (a) Temperature dependence of the FMR frequency. The inset highlights the temperature dependence of FMR for the sample B. (b) FMR linewidth vs. temperature for samples A-C. All data are measured in the external field of 20mT. Arrows indicate the corresponding anomalies observed in the M(T) curve.

FIG. 3. (Color on-line) (a) Magnetic field dependences of magnetization measured for the sample C in the proximity to 60K where an anomaly in the high field magnetization is observed. P and AP denote parallel and antiparallel alignment of free-pinned layers respectively.(b) Temperature dependence of the soft layer switching field for the three samples studied. (c) Temperature dependence of the $M^*_S$ (T) normalized by $M^*_S$ (5K). The inset highlights dependence of $M^*_S$ (T)/$M^*_S$ (5K) for the single Py film.

FIG. 4. (Color on-line) Top: TEM images of the single Py film and one of the MTJs (B). The dashed lines indicate the insulating barrier profile which shows presence of regions with anticorrelated roughness (marked with dotted arrows). Bottom: Sketch explaining proposed magnetization configuration and the energy profiles of both soft and hard layers in the regions with anticorrelated roughness. Relative and absolute minima correspond to out-of-plane and in-plane magnetizations respectively while spring indicates soft/hard layer coupling for: (a) T«$T_R$. (b) T≤$T_R$. (c) T≥$T_R$ and (d) T»$T_R$ conditions. Dotted lines indicate stray fields. Dashed arrows show non-equilibrium magnetization.

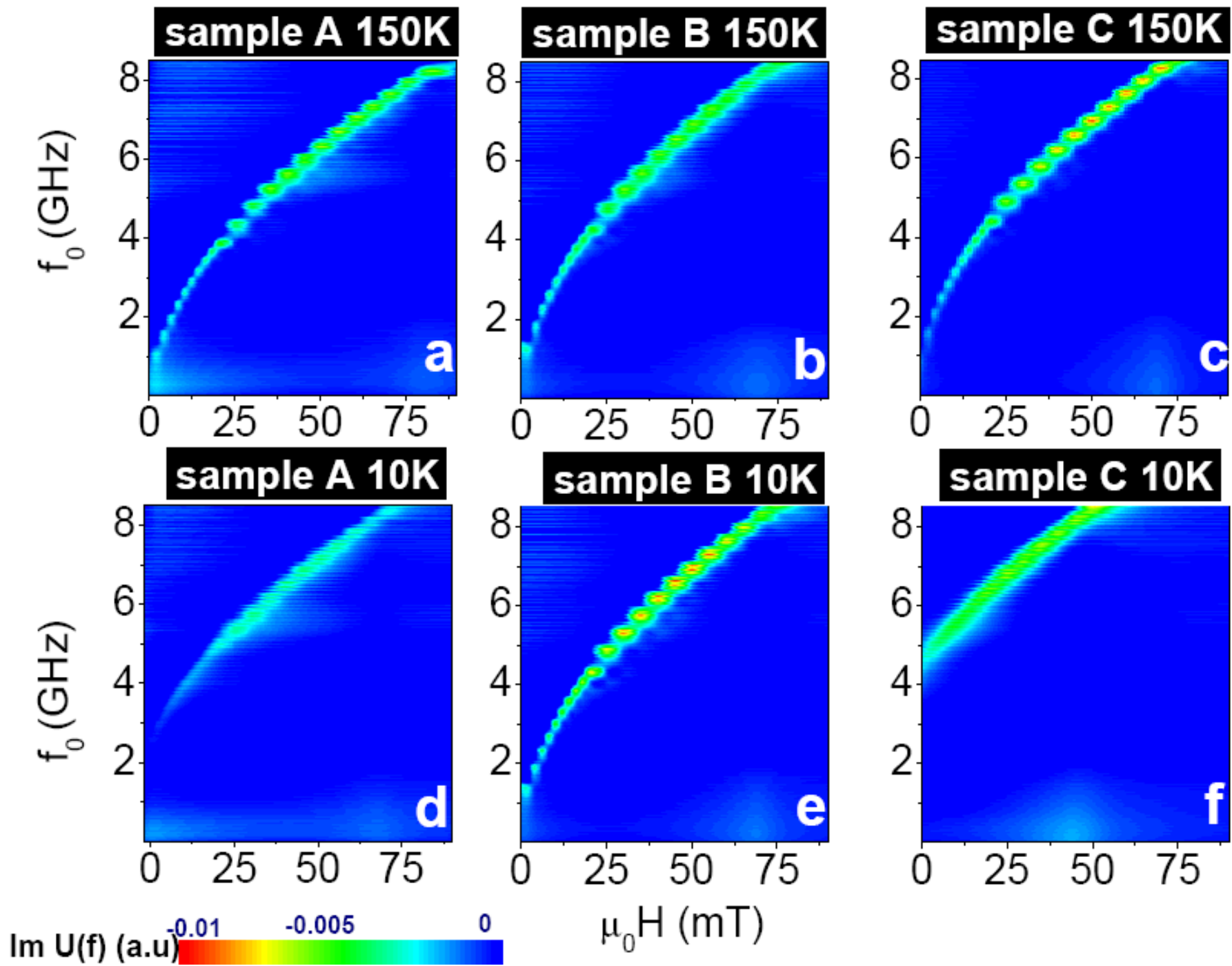

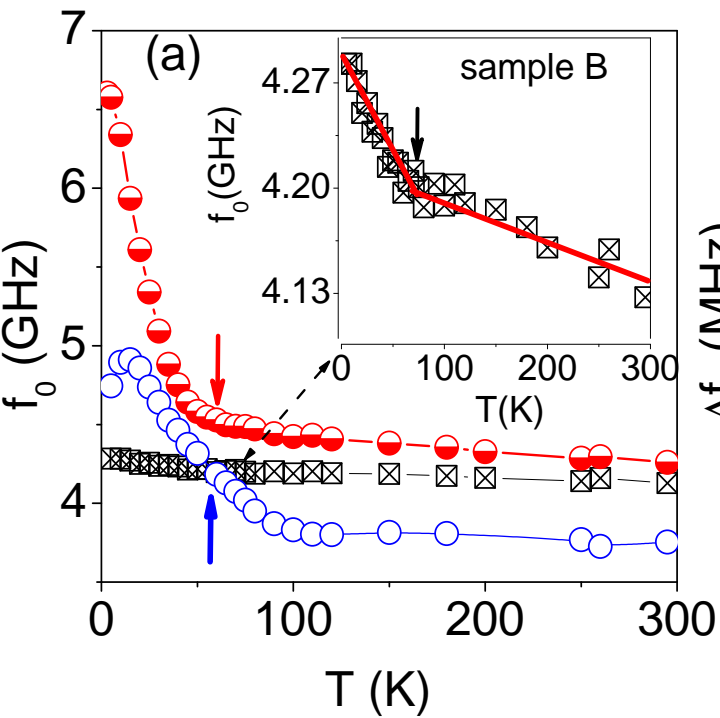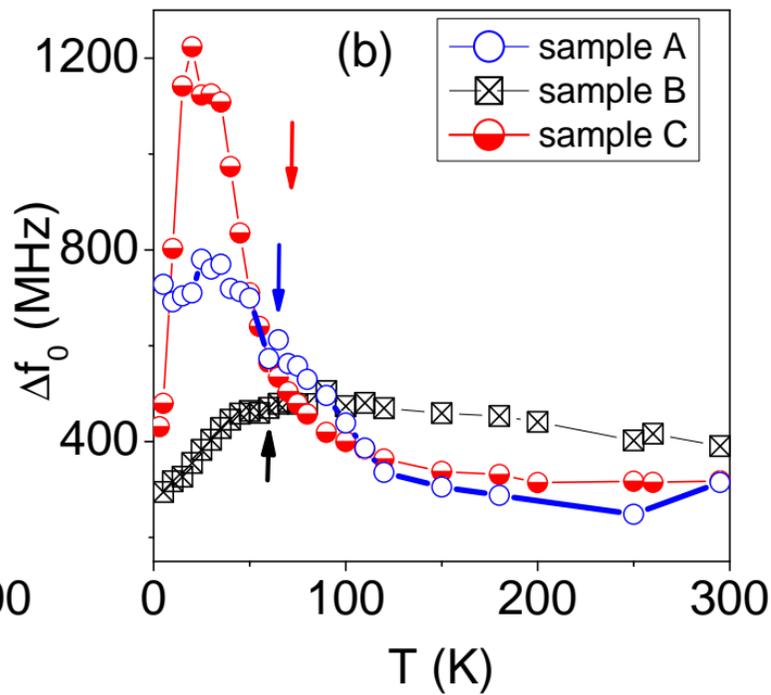

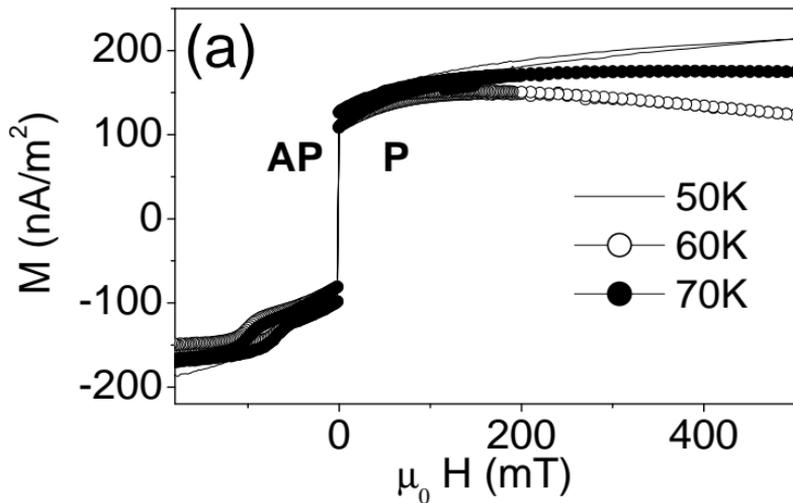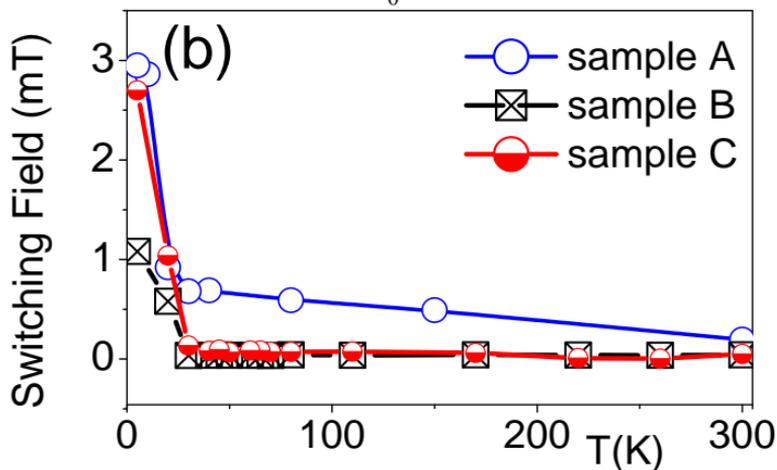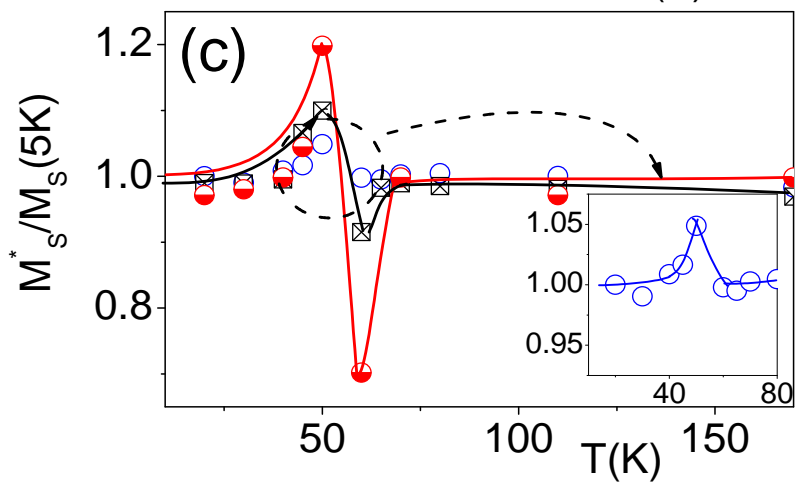

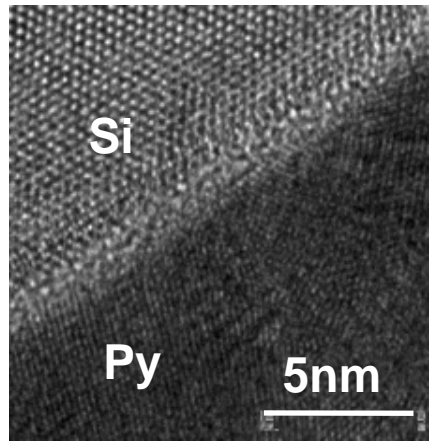
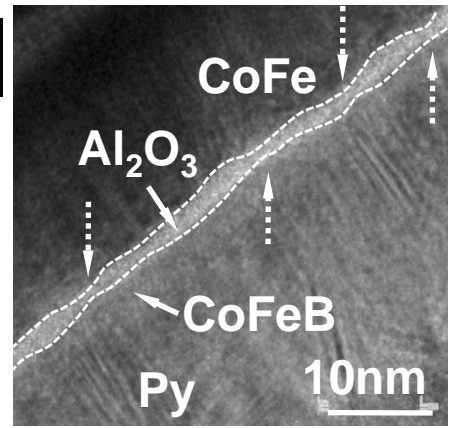
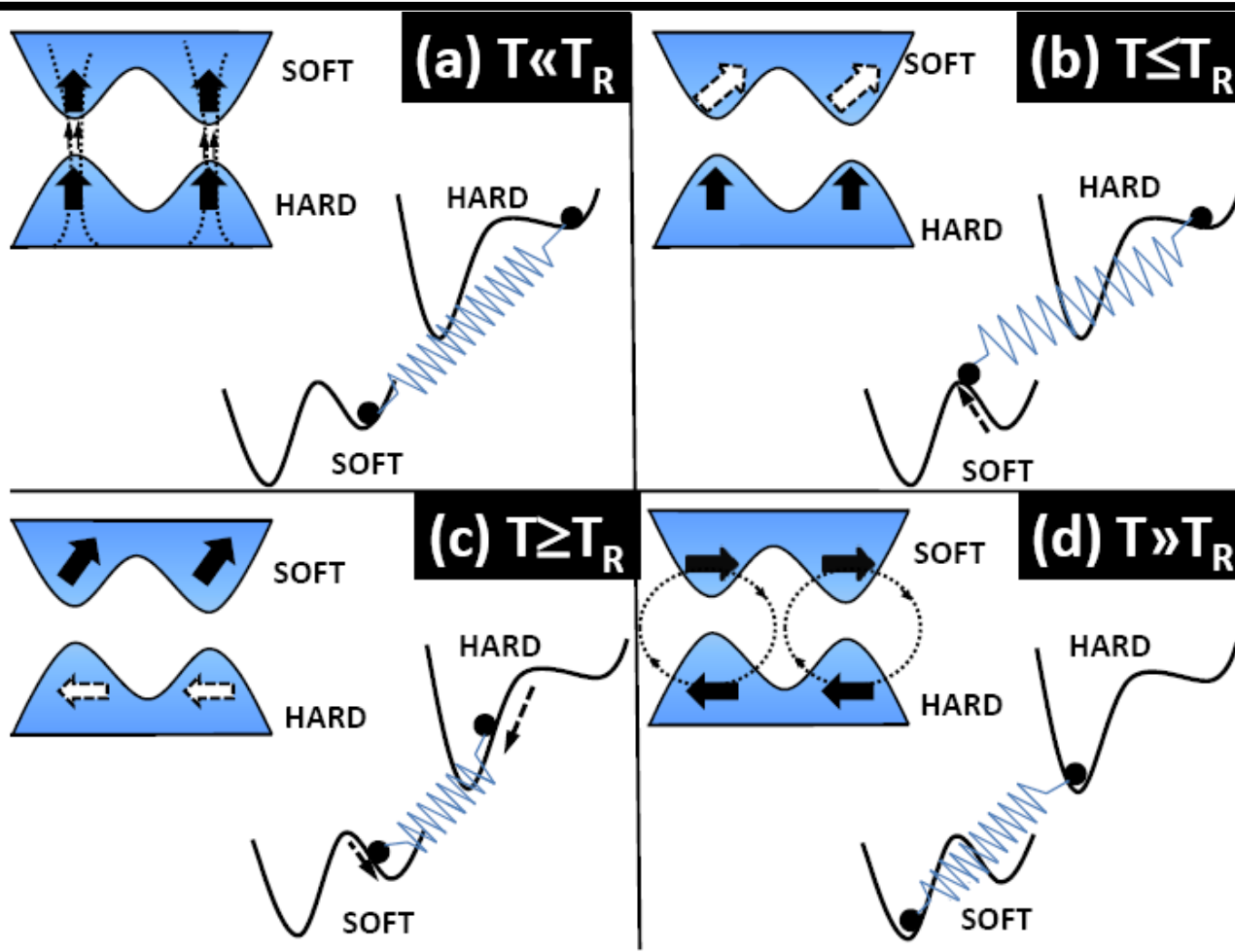